\documentclass[a4paper,fleqn]{article}
\usepackage{amsmath,amssymb,amsfonts,latexsym}

\usepackage{graphicx}
\usepackage[colorlinks,linkcolor=blue,anchorcolor=black,citecolor=red]{hyperref}
\usepackage{color}
\usepackage[section]{placeins}
\usepackage{subfigure}
\usepackage{bm}
\usepackage{enumerate}
\usepackage{cite}
\usepackage{caption}
\newcommand{\N}{{\mathbb{N}}}  
\newcommand{\R}{{\mathbb{R}}}  

\newcommand{\feta}{\bm{\eta}}

\newcommand{\be}{\begin{equation}}
\newcommand{\ee}{\end{equation}}
\newcommand{\bee}{\begin{equation*}}
\newcommand{\eee}{\end{equation*}}

\begin{document}

\title{Dynamics of condensation in the totally asymmetric inclusion process}
\author{Jiarui Cao, Paul Chleboun, Stefan Grosskinsky\\[2mm]
{\small Mathematics Institute, University of Warwick, Coventry CV4 7AL, UK}}
\date{\today}                                           

\maketitle
\begin{abstract}
We study the dynamics of condensation of the inclusion process on a one-dimensional periodic lattice in the thermodynamic limit, generalising recent results on finite lattices for symmetric dynamics. Our main focus is on totally asymmetric dynamics which have not been studied before, and which we also compare to exact solutions for symmetric systems. We identify all relevant dynamical regimes and corresponding time scales as a function of the system size, including a coarsening regime where clusters move on the lattice and exchange particles, leading to a growing average cluster size. Suitable observables exhibit a power law scaling in this regime before they saturate to stationarity following an exponential decay depending on the system size. Our results are based on heuristic derivations and exact computations for symmetric systems, and are supported by detailed simulation data.
\end{abstract}

\section{Introduction}\label{sec:intro}

The inclusion process is a continuous-time stochastic particle system where particles perform independent random walks on a lattice and, in addition, interact via an attractive inclusion mechanism. The rates of the latter are proportional to the product of occupation numbers of departure and arrival sites. The process was first introduced in \cite{Giardina:2007em} as a dual of a model of heat conduction and has been further developed as a bosonic counterpart of the exclusion process in \cite{Giardina:2010jf}. It has been shown that the system can exhibit a condensation transition in the limit of vanishing diffusion parameter $d$, which encodes the rate of independent motion of the particles. The strong inclusion interaction leads to typical stationary configurations where a single lattice site contains almost all the particles in the system. This has been established rigorously on finite lattices in \cite{Grosskinsky:2011uh} and in the thermodynamic limit in \cite{Chleboun:2012vc}. Besides applications to energy transport \cite{Giardina:2009jx}, the inclusion process can also be interpreted as a multi-allele version of the Moran model \cite{MORAN:1962uk}. In this case $d$ can be interpreted as the mutation rate, which is typically very small with respect to the reproductive time scale (see \cite{Arnoldt:2012bo} and references therein), and condensation corresponds to fixation of a particular species. There is also a different model \cite{Reuveni:2012cv, Reuveni:2013vs} that has been named inclusion process, where whole clusters of particles can jump simultaneously, which we do not consider in this paper.

In addition to characterising the stationary properties of condensation in stochastic particle systems, understanding the dynamics of condensation poses a very natural and interesting problem. For the symmetric inclusion process, the dynamics of the condensate formation and subsequent motion have been studied rigorously in \cite{Grosskinsky:2013ji} in the limit of infinitely many particles on a fixed, finite lattice. In this paper we extend these results in a non-rigorous way to spatially homogeneous, asymmetric systems in the thermodynamic limit, i.e. diverging lattice size with a finite particle density. For simplicity, we focus on the totally asymmetric system in one dimension with periodic boundary conditions, and also discuss some aspects of symmetric systesm for comparison. We identify and describe in detail various regimes of the condensation dynamics, including most importantly a coarsening regime where particle clusters move and exchange particles, following a power law scaling. We also describe the exponential approach to equilibrium in the saturation regime, and the initial nucleation dynamics where isolated particle clusters form on a fast time scale. 

The coarsening behaviour in condensing systems has already been studied heuristically in \cite{Drouffe:1999jk} and subsequent work for zero-range processes \cite{Evans:2000vw,Godreche:2003gb,Grosskinsky:2003eq,Godreche:2005kj,Evans:2005fz} and related models \cite{Godreche:2001vo,Godreche:2007gx, Ferretti:2008ij}. There is also a significant literature on the dynamics of condensation in spatially heterogeneous models (see \cite{Godreche:2012ky} and references therein). Until recently, mathematically rigorous results have been restricted to stationary measures (see \cite{Jeon:2000wq,Grosskinsky:2003eq,Armendariz:2008cp} and references therein), and the dynamics of condensation continue to attract significant research interest. The metastable condensate dynamics for zero-range processes have been derived rigorously in a series of papers \cite{Beltran:2010cm,Beltran:2011cg,Beltran:2012be,Beltran:2013tz}, and the hydrodynamic behaviour has been studied heuristically in \cite{Schutz:2007ec}. A rigorous description of the coarsening dynamics are currently under investigation \cite{Jara:2013ta}. In contrast to zero-range models, in the inclusion process and related models condensates are mobile on the coarsening time scale and coarsening is in fact driven by condensate motion and interaction \cite{Waclaw:2012ww,Grosskinsky:2013ji, Chleboun:2013um}. 
Further developments in this direction include explosive condensation in a totally asymmetric model \cite{Waclaw:2012ww} which exhibits a slinky motion of the condensate, also observed recently in \cite{Hirschberg:2012ee} for non-Markovian zero-range dynamics. In this paper we are able to give a detailed picture of this phenomenon in the asymmetric inclusion process by studying the interaction of two clusters. Further recent results on non-condensing inclusion processes include a hydrodynamic scaling limit for the symmetric system making use of self-duality of the model \cite{Opoku:2013tb}, which is not available for the asymmetric model we consider in this paper.

The rest of the paper is organised as follows. In Section \ref{sec:ip} we introduce the model, and describe the condensation transition and the different regimes of the dynamics. In Section \ref{sec:nucl} we describe the nucleation regime of cluster formation starting with a closed form solution for the symmetric system and analogous numerical observations for the asymmetric case. In Section \ref{sec:cond} we describe the mechanisms of cluster motion and interaction, which form the basis of the coarsening and saturation dynamics at larger time scales discussed in Section \ref{sec:cors}. We give heuristic derivations of scaling laws for symmetric and asymmetric systems, and support our predictions with detailed simulation data. In the appendix we provide further details on the nucleation regime in the asymmetric case, a theoretical description of which remains an intriguing problem but is not the main focus of this paper.

\section{Inclusion process}\label{sec:ip}

\subsection{Definition and notation}\label{sec:ip:def}

The inclusion process $(\bm{\eta}(t):t\geq 0)$ is a stochastic particle system defined on a lattice $\Lambda_{L}$ of $L$ sites, which we fix to be one-dimensional with periodic boundary conditions. Configurations are denoted by $\bm{\eta}=(\eta_{x}:x\in\Lambda_{L})\in X_{L}$ where $\eta_{x}\in\N$ is the number of particles at site $x\in\Lambda_{L}$. This can be arbitrarily large and the full state space is given by $X_{L}=\N^{\Lambda_{L}}$. 

The dynamics are defined by the generator acting on bounded test functions $f:X_{L}\mapsto \R$,
\be
\label{generator}
\mathcal{L}_{L}f(\bm{\eta})=\sum_{x,y\in\Lambda_{L}}p(x,y)\eta_{x}(d+\eta_{y})(f(\bm{\eta}^{x,y})-f(\bm{\eta})) \, ,
\ee
where $\bm{\eta}^{x,y}$ is the configuration after moving one particle from site $x$ to site $y$, i.e. $\eta_{z}^{x,y}=\eta_{z}-\delta_{z,x}+\delta_{z,y}$. The parameter $d_{L}$ scales with the system size, and determines the relative rate of the independent random walk of particles in comparison to the interacting inclusion part given by the product $\eta_{x}\eta_{y}$. The $p(x,y)\geq0$ are transition rates of an arbitrary, irreducible random walk on $\Lambda_{L}$. In this paper we focus on nearest-neighbour symmetric (SIP) and totally asymmetric (TASIP) dynamics, where $p(x,y)=1/2(\delta_{y,x+1}+\delta_{y,x-1})$ or $p(x,y)=\delta_{y,x+1}$, respectively. Details of inclusion processes on more general lattices including open boundaries can be found in \cite{Giardina:2009jx, Giardina:2010jf, Grosskinsky:2011uh}.

\subsection{Stationary distributions}\label{sec:ip:statdis}

Stationary product measures for the SIP were identified in \cite{Giardina:2007em,Giardina:2009jx} and extended in \cite{Chleboun:2013um,Grosskinsky:2011uh} to more general spatial rates $p(x,y)$, including the totally asymmetric case. Since we focus on translation invariant systems, we have homogeneous product measures 
\bee
\nu_{\phi}^{L}[d\feta]=\prod_{x\in\Lambda_{L}}\bar{\nu}_{\phi}(\eta_{x})d\feta\, \quad \,\,\, \textrm{with} \,\,\, \bar{\nu}_{\phi}(n)=\frac{1}{z(\phi)}w(n)\phi^{n} \,\, ,
\eee
given by a product density $\nu^{L}_{\phi}$ with respect to product counting measure $d\feta$. $\phi\geq0$ is the fugacity parameter controlling the particle density. The stationary weight is given by 
\bee
w(n)=\frac{\Gamma(d_{L}+n)}{n!\Gamma(d_{L})} \, ,
\eee
and the single-site partition function by
\be
z(\phi)=\sum_{k=0}^{\infty}w(k)\phi^{k}=(1-\phi)^{-d_{L}}\, .
\label{partition}
\ee
Since the partition function diverges as $\phi\nearrow 1$, the measures exist for all $\phi\in[0,1)$ and constitute the grand canonical ensemble \cite{Grosskinsky:2011uh,Chleboun:2013um}. The average particle density is a function of $\phi$, and is given by 
\be
R(\phi):=\mathbb{E}_{\phi}[\eta_{x}]=\sum_{k=0}^{\infty}k\bar{\nu}_{\phi}(k)=\phi\partial_{\phi}\log_{z}(\phi)=\frac{d_{L}\phi}{1-\phi} \, .
\label{avgden}
\ee

For the TASIP the average stationary current per site is given by the average jump rate off a site, which also determines the corresponding diffusivity for the symmetric system. Under the grand canonical ensemble this is given by 
\bee
j^{gc}(\phi)=\mathbb{E}_{\phi}[\eta_{x}(d_{L}+\eta_{x+1})]=R(\phi)(R(\phi)+d_{L}) \, ,
\eee
depending only on the particle density and $d_{L}$.

Since the total number of particles is conserved and is the only conserved quantity, the inclusion process (\ref{generator}) is irreducible on the finite subsets $X_{L,N}=\left\{\feta\in X \, : \, \sum_{x=1}^{L}\eta_{x}=N \right\}$ with unique stationary measure $\pi_{L,N}$. This family of measures for $N\in\N$ constitutes the canonical ensemble, and is extremal for all $L$. It can be written in conditional form 
\bee
\pi_{L,N} =\nu_{\phi}\left[\, .\,  \left|   \,\sum_{x=1}^{L}\eta_{x}=N\right]\right.\ ,
\eee
which is independent of the choice of $\phi$. 

Note that we do not address infinite lattices directly in this paper, which is another interesting problem for which the definition of asymmetric dynamics of the process poses the first challenge. While symmetric processes can be defined via duality (cf.~\cite{Carinci:2012wh}), for asymmetric dynamics there is so far no result that establishes a well defined time evolution and rules out a divergence of particle flux from infinity in finite time.

\subsection{Condensation }\label{sec:ip:cond}

For fixed $L$ and $d_L$ the range of densities is $R([0,1))=[0,\infty )$ and the process does not exhibit condensation in the usual sense of zero-range processes \cite{Evans:2000vw, Grosskinsky:2003eq} or related models \cite{Chleboun:2013um}, where this range is bounded. But it has been established in \cite{Grosskinsky:2011uh,Chleboun:2012vc} that in the thermodynamic limit with vanishing diffusion rate
\be\label{tlim}
L,N \rightarrow \infty\, ,\ d_L \to 0\quad\textrm{such that}\quad\frac{N}{L}\rightarrow \rho >0\quad\mbox{and}\quad d_L L\to 0\ ,
\ee
the system exhibits complete condensation. 
In this case,
\be
\max_{x\in\Lambda} \eta_x /N\to 1\quad \mbox{in distribution }\pi_{L,N}\ ,
\ee
so if the diffusion rate scales as $d_L \ll 1/L$ almost all particles in the system condense on a single site. Furthermore, in \cite{Chleboun:2012vc} stationary large deviations for the maximum occupation number are computed in the limit (\ref{tlim}), and for condensing systems the most likely value scales as the total number of particles $N$ in the system. 
We will assume $d_{L} \ll 1/L$ for the rest of the paper and for all simulation results presented we use $d_{L}=\frac{1}{L^{2}}$, but have checked the validity also for other scalings of $d_L$.

In contrast to zero-range processes, the condensate and large clusters move on the same time scale as the system equilibrates. Motion and interaction of clusters dominates the coarsening process, as will be explained in detail in the following. This has been established rigorously in \cite{Grosskinsky:2013ji} for the simpler setting of symmetric systems on fixed lattices.

To describe the dynamics of condensation we consider the second moment
\be\label{smom}
\sigma^2 (t)= \mathbb{E} \left[\eta_{x}^{2} (t)\right]\quad\mbox{for some }x\in\Lambda_L \ ,
\ee
which does not depend on $x$ since we will assume the initial distribution to be translation invariant. This is the simplest observable that captures the temporal evolution of the condensed phase, since the first moment is constant in time due to conservation of the number of particles. 
Due to spatial homogeneity, in simulations we measure $\sigma^{2}(t)$ by spatial average $\left\langle 1/L\sum_{x=1}^{L}\eta_{x}^{2} \right\rangle$ to have better statistics, where $\langle\cdot\rangle$ denotes averaging over a large number (typically 200 in our simulations) of realisations.

\begin{figure}
\centering
\subfigure{
\includegraphics[height=6cm]{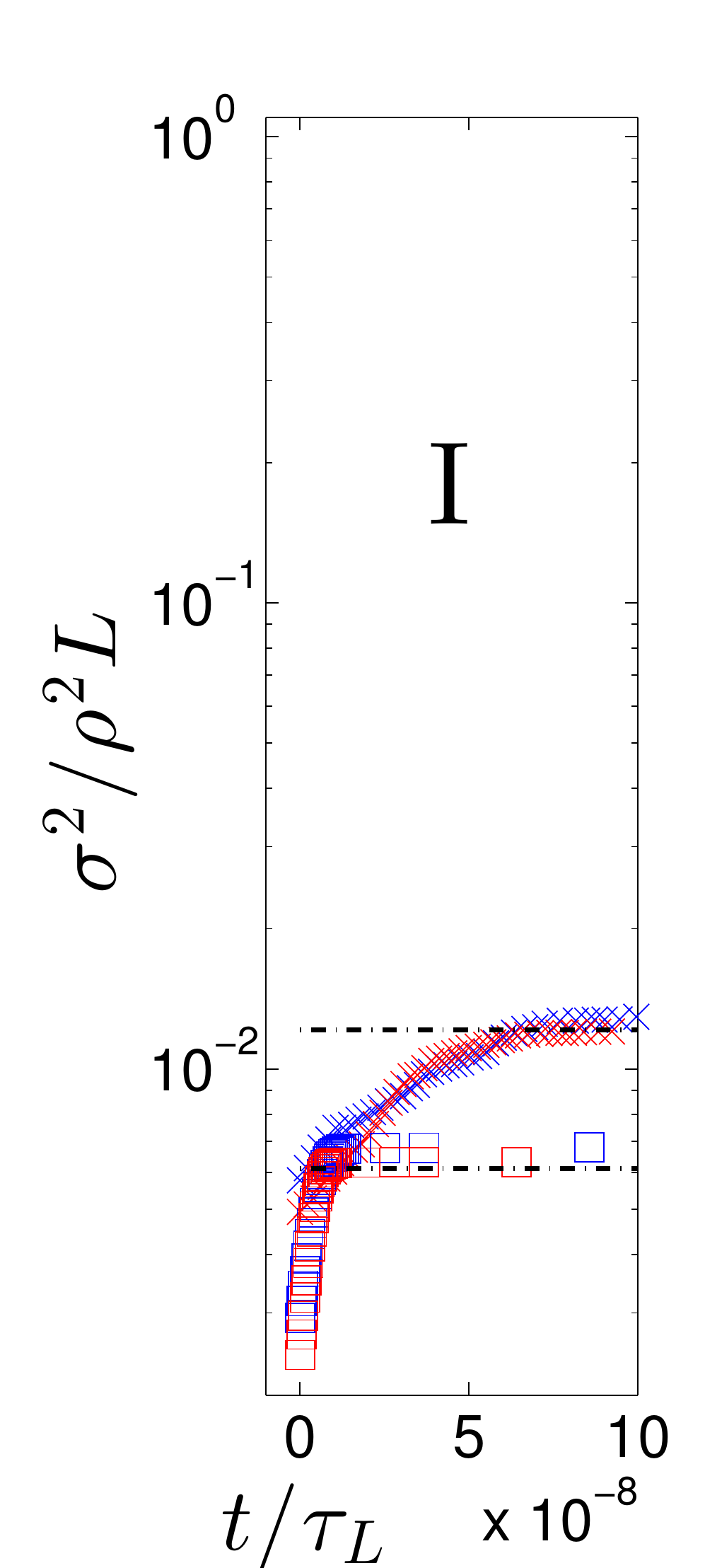}}
\subfigure{
\includegraphics[height=6cm]{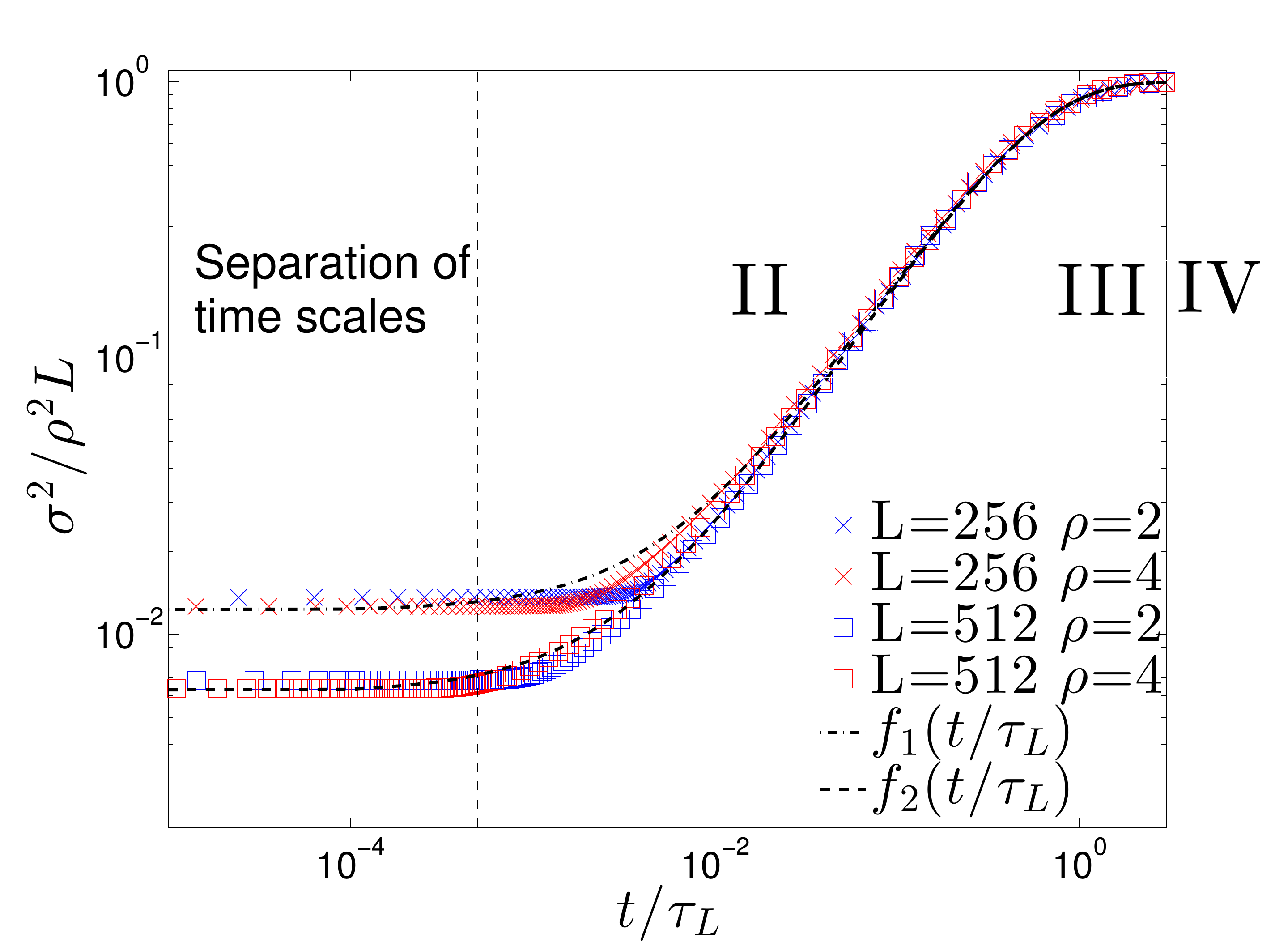}}
\caption{Illustration of different dynamical regimes in TASIP. The rescaled observable $\sigma^2 (t)/\rho^{2}L$ (\ref{smom}) is shown against rescaled time $t/\tau_{L}$ with $\tau_L =L/d_L$ (cf. (\ref{TimeScaleTASIP})) and $d_L=1/L^2$. $f_{1}$ and $f_{2}$ are exponential functions (cf. (\ref{exlaw})) describing the long-term asymptotics, with initial values $\sigma^{2}_0$ fitted to data for $L=256, \rho=4$ and $L=512, \rho=4$, respectively. $\sigma^{2}(0)$ is calculated at the end of the fast nucleation process explained in Section \ref{sec:nucl}. Data points are averaged over 200 realisations, errors are bounded by the size of the symbols.}
\label{TASIP2ndMoment}
\end{figure}

We consider canonical initial conditions where $N$ particles are placed uniformly and independently on the lattice, which leads to $\feta (0)$ having a symmetric multinomial distribution with $N$ trials and success probability $1/L$. For $L\rightarrow\infty$ and $N/L\rightarrow\rho$ the occupation numbers are asymptotically independent Poisson random variables $\eta_{x} (0)\sim Poi(\rho)$, and have second moment $\sigma^{2} (0)=\rho(1+\rho)$. Furthermore, in stationarity as $t\to\infty$ we know that up to fluctuations all particles condense on a single site, and we expect $\sigma^{2}_{L} (t)\approx \frac{1}{L}(\rho L)^{2}=\rho^{2}L$. So we consider the rescaled variable $\sigma^{2}_{L} (t)/\rho^{2}L$, which increases from very small values of order $1/L$ to $1$ during the formation of the condensate from homogeneous initial conditions. This process can be divided into four different regimes (see Figure \ref{TASIP2ndMoment}):
\\
\\
\begin{enumerate}[(I).]
\item{\bf Nucleation Regime}: Due to the inclusion rate $\eta_{x}\eta_{y}$, neighboring pairs of sites exchange particles with order $1$ rates until the process reaches a state where all occupied sites are separated by at least one empty site. This happens simultaneously everywhere on the lattice and takes at most of order $\log L$ time. 
After this regime, a fraction of at most $1/2$ of all sites is occupied and particles can only jump to another site by the diffusion part of the dynamics with slow rate $d_L$. Details can be found in Section \ref{sec:nucl}.

\item {\bf Coarsening Regime}: Particle clusters formed in regime (I) can move to empty neighbouring sites or exchange particles at rate $\eta_{x}d_{L}$, but typically do not split on this timescale. This drives a coarsening process with a decreasing number of clusters of increasing size, which grow to clusters of order $N$ size. This coarsening process happens on a characteristic time scale $1/d_{L}$, as explained in detail in Section \ref{sec:cors}. As expected, $\sigma^2 (t)$ follows an approximate power law in this regime.

\item {\bf Saturation Regime}: The coarsening scaling law no longer holds since the system reaches its finite size limit, and the remaining clusters merge to form a single condensate. As expected close to stationarity, the observable $\sigma^{2} (t)$ converges exponentially to its stationary value, as explained in detail in Section \ref{sec:cors:ex}. The characteristic time scale for this regime is up to constant factors the relaxation time of the system, and turns out to be of order $\tau_{L} =L/d_L$ for the TASIP and $L^2 /d_L$ for the SIP (see Section \ref{sec:cond:timescale}). 

\item{\bf Stationary Regime}: Once there is only a single condensate left on the lattice, it continues to move according to the same rules and time scales as in regimes II and III. The observable $\sigma^2$ does not detect this motion, but it can be studied by defining the location of the maximum occupation number as relevant observable as has been done on fixed lattices in \cite{Grosskinsky:2013ji}, or in \cite{Beltran:2011cg} for zero-range processes. 

\end{enumerate}

In Figure \ref{TASIP2ndMoment} we illustrate the condensation dynamics on the total relaxation time scale $\tau_{L}$. 
Details of the time scale will be discussed in Section \ref{sec:cond:timescale}. As the nucleation regime occurs on a time scale of at most $\log L$, it finishes immediately and just determines the initial condition for the coarsening regime. Note that the exponential approximation for the saturation regime also fits the data in the coarsening regime very well. This is a peculiarity due to the linear coarsening law for the TASIP as explained in Section \ref{sec:cors}, and does not hold for the SIP or in general.

\section{Nucleation regime}\label{sec:nucl}
The nucleation regime starts with the initial distribution of particles, which we take to be a uniform multinomial for simplicity. It ends when no particles reside on successive sites which can be defined by the hitting time
\be\label{nuctime}
T:= \inf\left\{t\geq 0 : \sum_{x\in_{\Lambda_{L}}}\eta_{x}(t)\eta_{x+1}(t)=0\right\} \ .
\ee
Under our condition of weak diffusion $d_L \ll 1/L$, the inclusion effect completely dominates this regime and the time scale $\mathbb{E} [T]$ turns out to scale as $\log L$, which is much faster compared to all other regimes. We will take two different approaches for the TASIP and the SIP, starting with the simpler symmetric case.

\subsection{Symmetric case}\label{sec:nucl:sip}
In the SIP we can derive closed relations for the dynamics of correlation functions due to symmetry. We consider the nearest neighbour product
\be\label{c1tdef}
c(1,t):=\mathbb{E}[\eta_{x}\eta_{x+1}]\quad \textrm{for some} \,\, x\in\Lambda_{L} \, ,
\ee 
since the observable $\eta_{x}\eta_{x+1}$ vanishes at the end of the saturation regime. Similar to $\sigma^{2}(t)$, $c(1,t)$ is also $x$-independent due to translation invariance and in simulations we measure $c(1,t)$ by the spatial average $\big\langle 1/L\sum_{x=1}^{L}\eta_{x}\eta_{x+1}\big\rangle$ as described earlier. With our initial conditions we have $c(1,0)=\rho^{2}$ for large $L$, and $c(1,t)\rightarrow 0$ with increasing time. Applying the generator (\ref{generator}) to the test function $f(\bm{\eta})=\eta_{x}\eta_{x+1}$ for some $x\in\Lambda_{L}$, we get

\begin{eqnarray*}
\mathcal{L}(\eta_{x}\eta_{x+1})&{=}&\frac{1}{2}\eta_{x-1}(d_{L}+\eta_{x})[(\eta_{x}+1)\eta_{x+1}-\eta_{x}\eta_{x+1}]\\
&{+}&\frac{1}{2}\eta_{x}(d_{L}+\eta_{x-1})[(\eta_{x}-1)\eta_{x+1}-\eta_{x}\eta_{x+1}]\\
&{+}&\frac{1}{2}\eta_{x}(d_{L}+\eta_{x+1})[(\eta_{x}-1)(\eta_{x+1}+1)-\eta_{x}\eta_{x+1}]\\
&{+}&\frac{1}{2}\eta_{x+1}(d_{L}+\eta_{x})[(\eta_{x}+1)(\eta_{x+1}-1)-\eta_{x}\eta_{x+1}]\\
&{+}&\frac{1}{2}\eta_{x+1}(d_{L}+\eta_{x+2})[\eta_{x}(\eta_{x+1}-1)-\eta_{x}\eta_{x+1}]\\
&{+}&\frac{1}{2}\eta_{x+2}(d_{L}+\eta_{x+1})[\eta_{x}(\eta_{x+1}+1)-\eta_{x}\eta_{x+1}]\\
&{=}&\!\!\!-\eta_{x}\eta_{x+1} {+} \frac{1}{2}d_{L}(-4\eta_{x}\eta_{x+1}{+}\eta_{x-1}\eta_{x+1}{+}\eta_{x}\eta_{x+2}{+}\eta_{x}^{2}{+}\eta_{x+1}^{2}{-}\eta_{x}{-}\eta_{x+1})\, .
\end{eqnarray*}

\begin{figure} 
\centering
  \subfigure[]{
    \begin{minipage}[b]{0.5\textwidth} 
      \centering 
      \includegraphics[height=4.5cm]{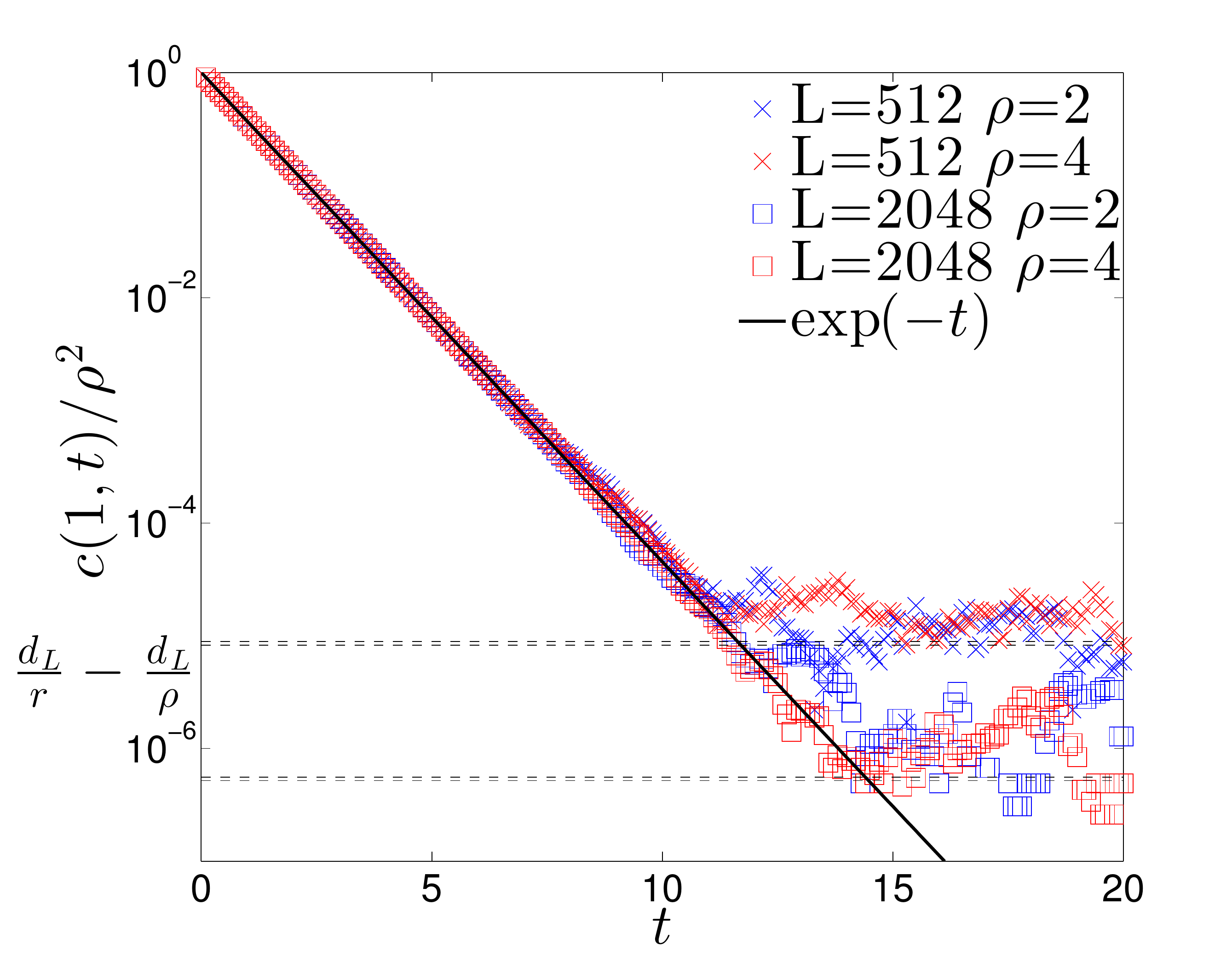} 
    \end{minipage}
    }%
  \subfigure[]{ 
    \begin{minipage}[b]{0.5\textwidth} 
      \centering 
      \includegraphics[height=4.5cm]{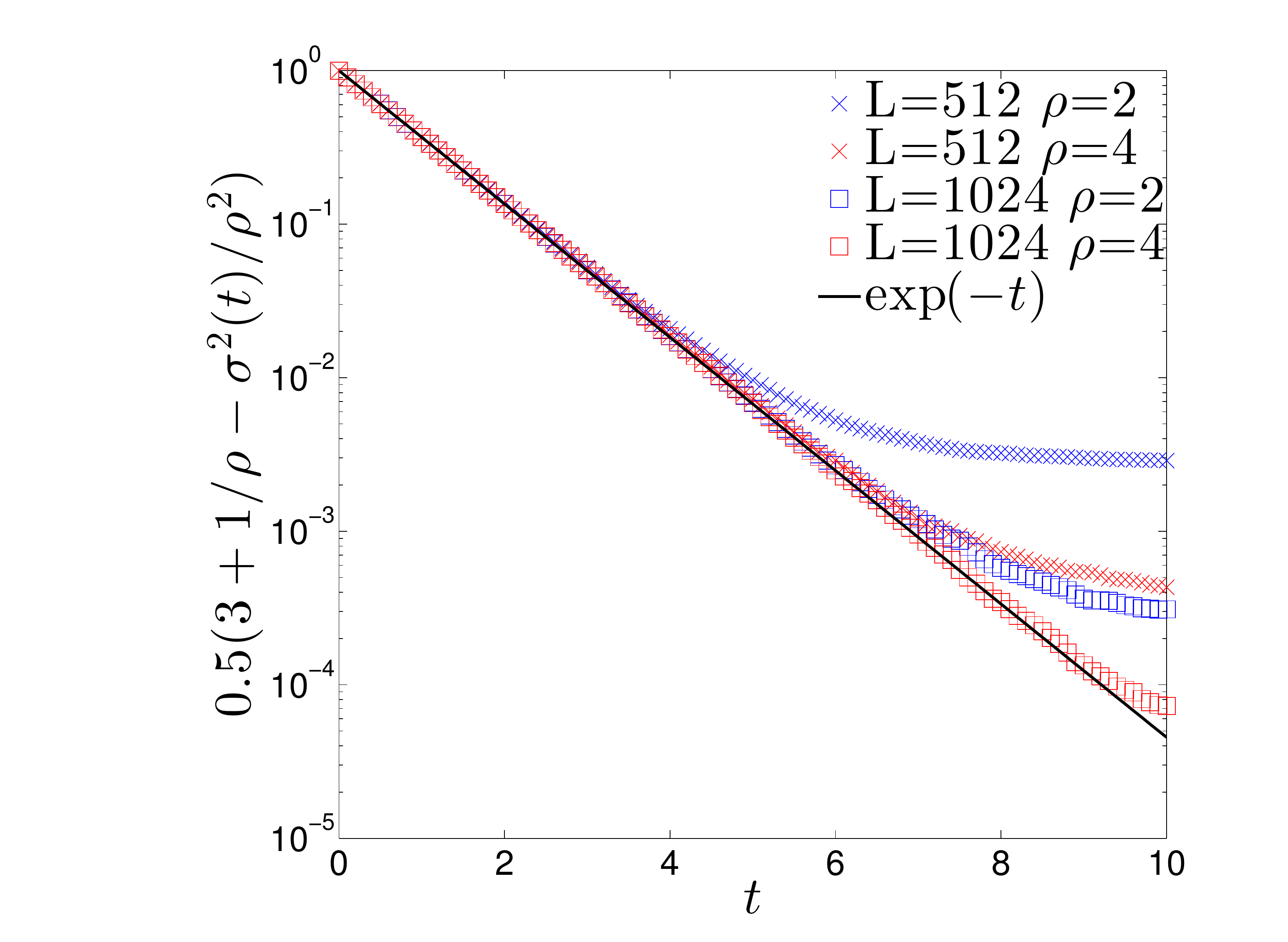} 
    \end{minipage}
    } 
 \caption{Exponential behavior of $c(1,t)$ (\ref{c1tdef}) and $\sigma^{2}(t)$ (\ref{smom}) for the SIP in the nucleation regime. (a) Exponential decay of $c(1,t)/\rho^{2}$ as given in (\ref{SIPProduct}). Dashed lines are fluctuation estimates $d_{L}/r-d_{L}/\rho$ for $L=512, 1024$, where we used numerical values for the ratio $r$ (\ref{Rdef}): $r|_{\rho=2}=0.3747$ and $r|_{\rho=4}=0.3865$.  (b) Exponential convergence of $\sigma^{2}(t)/\rho^{2}$ as given in (\ref{NRSIPSecMot}). The deviations for large time are determined by the finite size corrections in (\ref{NRFiCorr}). Data points are averaged over 2000 realisations. Errors are of the order $10^{-4}$.} 
  \label{SIPNR} 
\end{figure}

 In the nucleation regime all occupation numbers are of order $1$, 
so the second term in the last line is of order $d_{L}$ in expectation. Then by the standard evolution equation, 
\be
\frac{d}{dt}c(1,t)=\mathbb{E}[\mathcal{L}(\eta_{x}\eta_{x+1})]=-c(1,t)+O(d_{L}) \,.
\label{SIPNR_Ceq}
\ee
For large systems, $d_L \ll 1/L$ is negligible and we solve the above ODE with initial condition $c(1,0)=\rho^{2}$ to get simply
\be
c(1,t)=\rho^{2}e^{-t} \, .
\label{SIPProduct}
\ee
Figure \ref{SIPNR}(a) shows a data collapse for $c(1,t)$ confirming this prediction. The large time plateau is dominated by attempted motion of clusters onto empty neighbouring sites with slow rate $d_{L}$. This motion leads to temporary nearest-neighbour occupation and produces finite size fluctuations of the asymptotic values of $c(1,t)$, which vanish with increasing system size. Their size can be estimated, considering the contribution to $c(1,t)$ during the step of a cluster. We consider a time $t_{1}>\mathbb{E}[T]$ so that we expect to have reached the plateau in Figure \ref{SIPNR}(a). Then we can estimate $c(1,t_{1})$ by the following ergodic average with duration $T_{\textrm{step}}$
\be\label{statflu}
c(1,t_{1})\simeq\mathbb{E}\left[\int_{t_{1}}^{t_{1}+T_{step}}\eta_{x}(s)\eta_{x+1}(s)\,ds\bigg|\eta_{x}(t_{1})>0\right]\frac{\mathbb{P}[\eta_{x}(t_{1})>0]}{\mathbb{E}[T_{\textrm{step}}]} \,,
\ee
where $T_{\textrm{step}}$ is the random time for an attempted step of the cluster. It is not important if the cluster actually moves to site $y=x-1$ or $x+1$ or stays at $x$. As discussed in detail in Section \ref{sec:cond:isoclust}, $T_{\textrm{step}}$ is dominated by the slow rate to move the first particle, after which all remaining particles quickly follow due to the inclusion interaction, and we have $\mathbb{E}[T_{\textrm{step}}]\sim 1/(d_{L}m)$ where $m=\mathbb{E}[\eta_{x}(t_{1})|\eta_{x}(t_{1})>0]$ is the size of a typical cluster. On the other hand, the integral in the numerator vanishes for most of the time, and the expected holding time in an intermediate state $(\eta_{x},\eta_{x+1})=(k,m-k)$ for $k\in\{1,2,...,m-1\}$ is simply $1/k(m-k)$. The computation of $c(1,t_{1})$ (\ref{statflu}) reduces to a simple random walk problem as is described in Appendix \ref{Appendix2}. We get

\be
\mathbb{E}\left[\int_{t_{1}}^{t_{1}+T_{\textrm{step}}}\eta_{x}(s)\eta_{x+1}(s)\,ds \bigg|\eta_{x}(t_{1})>0\right] =\mathbb{E}\left[\sum_{k=1}^{K_{\textrm{step}}}\frac{k(m-k)}{k(m-k)}\right] =m-1\ ,
\ee
where we used that the expected number of steps $K_{\textrm{step}}$ of an excursion starting with $(\eta_{x},\eta_{x+1})=(1,m-1)$ is $\mathbb{E}[K_{\textrm{step}}]=m-1$ (cf~(\ref{excursion})). Denote the (random) fraction of occupied sites at the end of the nucleation regime at time $T$ (\ref{nuctime}) by

\be\label{Rdef}
R:=\frac{1}{L}\sum_{x\in\Lambda_{L}}\bm{1}_{\{\eta_{x}(T)>0\}} \, , \textrm{and its expectation by} \, r=\mathbb{E}[R] \,.
\ee
Therefore $\mathbb{P}[\eta_{x}(t_{1})>0]\simeq r$ and we get in (\ref{statflu})
\be\label{c1limit}
c(1,t_{1})\simeq r\frac{m-1}{1/(d_L m)}=rm(m-1) d_L =\rho^{2}d_{L}\left(\frac{1}{r}-\frac{1}{\rho}\right)
\ee
using also $m=\rho/r$ for the average size of a cluster. This is confirmed by dashed lines in Figure~\ref{SIPNR}(a). Note (\ref{c1limit}) only makes sense for $\rho>r$, but we are not interested in very small densities $\rho$ which affect the nucleation regime due to a large number of empty sites already in the initial configuration.

To understand the evolution of the second moment (\ref{smom}) we take the test function $f(\bm{\eta})=\eta_x^2$, for some $x\in\Lambda$. Similarly to the above computation we get
\be\label{NRFiCorr}
\mathcal{L}(\eta_{x}^{2}){=}\eta_{x}\eta_{x{+}1}{+}\eta_{x{-}1}\eta_{x}{+}d_{L}\!\! \left( \!\!{-}2\eta_{x}^{2}+\eta_{x}{+}\eta_{x}\eta_{x{-}1}{+}\eta_{x}\eta_{x{+}1}{+}\frac{1}{2}\eta_{x{-}1}{+}\frac{1}{2}\eta_{x{+}1}\!\!\right) .
\ee
Again, the terms of order $d_L$ are negligible for large $L$ and in expectation 
\be\label{SIPNR_C1}
\frac{d}{dt}\sigma^{2}(t)=\mathbb{E}[\mathcal{L}(\eta_{x}^{2}(t))]=2c(1,t)+O(d_{L})\, . 
\ee
Integrating with initial condition $\sigma^{2}(0)=\rho(1+\rho)$ we get 
\be
\sigma^{2}(t)=2\rho^{2}(1-e^{-t})+\rho^{2}+\rho \,\, .
\label{NRSIPSecMot}
\ee
Note that this leading order behaviour is independent of $L$ and converges
\be
\frac{\sigma^{2}(t)}{\rho^{2}} \rightarrow 3+\frac{1}{\rho}=\sigma_{0}^{2} \quad\mbox{as }t\to\infty\ .
\label{NRSIPSecMotLmt}
\ee
This is the value of $\sigma^{2}$ after the nucleation regime on large systems and therefore gives the initial value of coarsening regime $\sigma^2_0$ up to finite size corrections as confirmed in Figure \ref{Cors}.

\subsection{TASIP}\label{sec:nucl:tasip}

\begin{figure}
\begin{center}
\includegraphics[height=6cm]{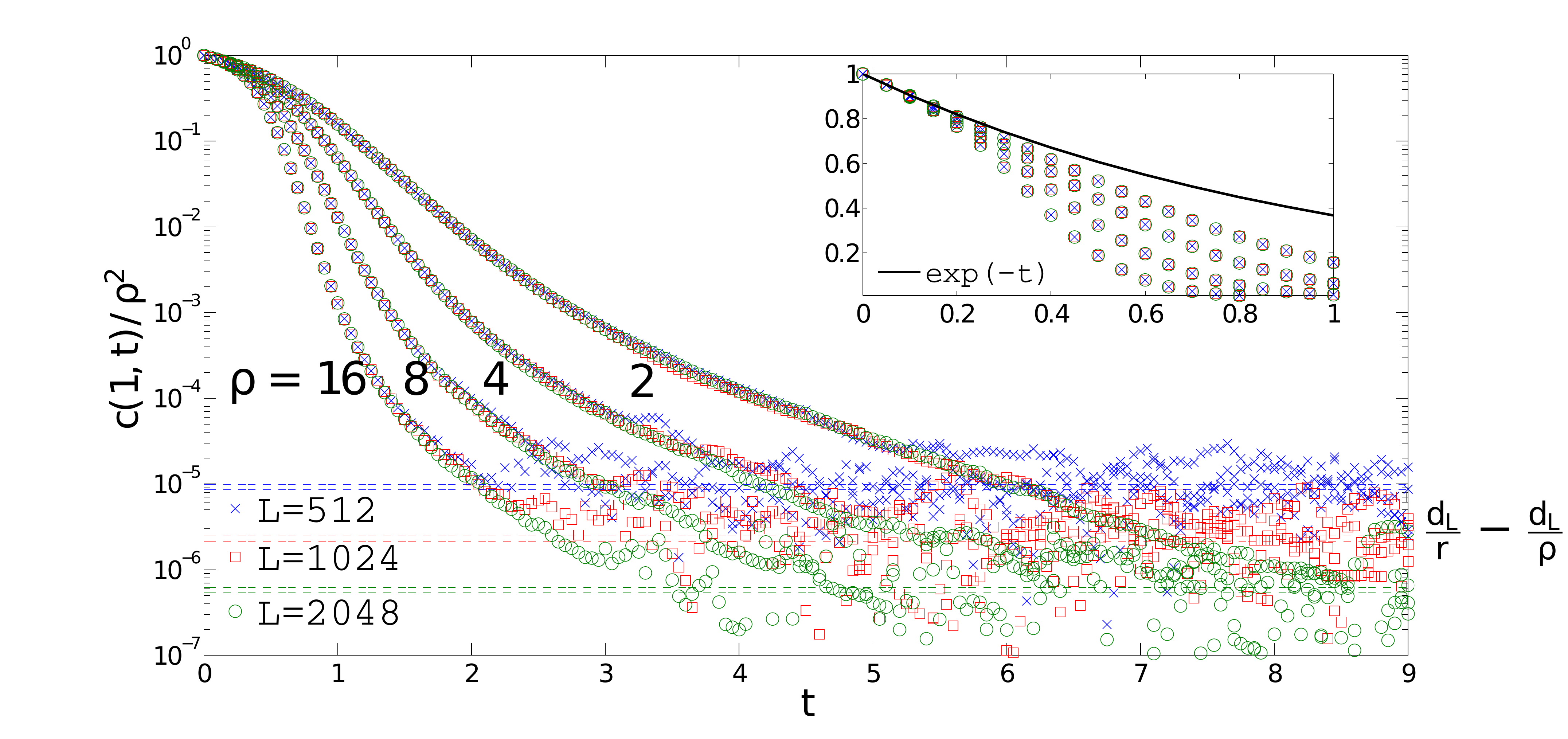}
\caption{ Super exponential decay of $c(1,t)$ for the TASIP in the nucleation regime. Dashed horizontal lines correspond to $L$-dependent corrections (cf (\ref{c1tasip})). For each system we give two lines by using the numerical maximal and minimal values of $r$ and $\rho$, where $r_{max}=0.4431$ for $\rho=16$ and $r_{min}=0.3850$ for $\rho=2$. The inset shows the initial behaviour which is approximately exponential. Data points are averaged over 2000 realisations. Errors are bounded by the size of the symbols until we observe the $L$-dependent corrections.}
\label{NRSIPCDecay}
\end{center}
\end{figure}

The reason we could get closed equations for correlation functions is related to self-duality of the SIP (see \cite{Carinci:2012wh} for more details). Due to the asymmetry, the TASIP is not self-dual and therefore the above technique does not lead to closed equations for $c(1,t)$ or other observables. In this subsection we will therefore focus mostly on simulations and approximations to understand the nucleation dynamics in TASIP. Applying the TASIP generator to the test function $f(\feta)=\eta_{x}\eta_{x+1}$ for some $x\in\Lambda_{L}$ we get analogously to the symmetric case

\begin{align*}
\mathcal{L}\left[\eta_{x}\eta_{x+1}\right] =& \eta_{x}\eta_{x+1}\left(-1+\eta_{x}-\eta_{x+1}+\eta_{x-1}-\eta_{x+2}\right)\\
+&d_{L}(\eta_{x-1}\eta_{x+1}-\eta_{x}\eta_{x+1}+\eta_{x}^{2}-\eta_{x}\eta_{x+1}-\eta_{x})\\
=&\eta_{x}\eta_{x+1}(\eta_{x-1}-\eta_{x+2}+\eta_{x}-\eta_{x+1}-1)+O(d_{L}) \, .
\end{align*}
Taking expectations and using translation invariance we get
\be
\frac{d}{dt}c(1,t)=-c(1,t)+\mathbb{E}\left[\eta_{x}\eta_{x+1}(\eta_{x}-\eta_{x+1})\right] \, .
\label{TASIPNR_Ceq}
\ee
This equation involves higher order correlation functions, and simple mean-field type arguments to close it fail to give reasonable predictions. In the nucleation regime interactions between clusters of particles are strong and complex, and correlations cannot be ignored. In fact, due to total asymmetry, given two neighbouring occupied sites the right one has higher occupation numbers on average and therefore the first order correction term in (\ref{TASIPNR_Ceq}) is negative, which leads to a super-exponential decay. For small times, dominated by initial conditions before correlations develop, the correction averages to zero and we observe an exponential decay as illustrated in Figure \ref{NRSIPCDecay}. The bulk decay shows a significant density dependence, but is independent of the system size $L$ for large enough systems. For large times, however, $c(1,t)/\rho^{2}$ converges to an $L$-dependent quasi-stationary value completely analogously to the symmetric case. Using the same arguments (see also Appendix \ref{Appendix2}) we get
\be\label{c1tasip}
c(1,t)\to\rho^{2}d_{L}\left(\frac{1}{r}-\frac{1}{\rho}\right) \quad \textrm{for large} \,\,\, t \, ,
\ee
 as confirmed by dashed lines in Figure~\ref{NRSIPCDecay}. Note that due to total asymmetry, the number of required particle moves for a cluster of $m$ particles to move one step on the lattice is precisely $m-1$, which simplifies the argument. 


\begin{figure}
\begin{center}
\includegraphics[height=6cm]{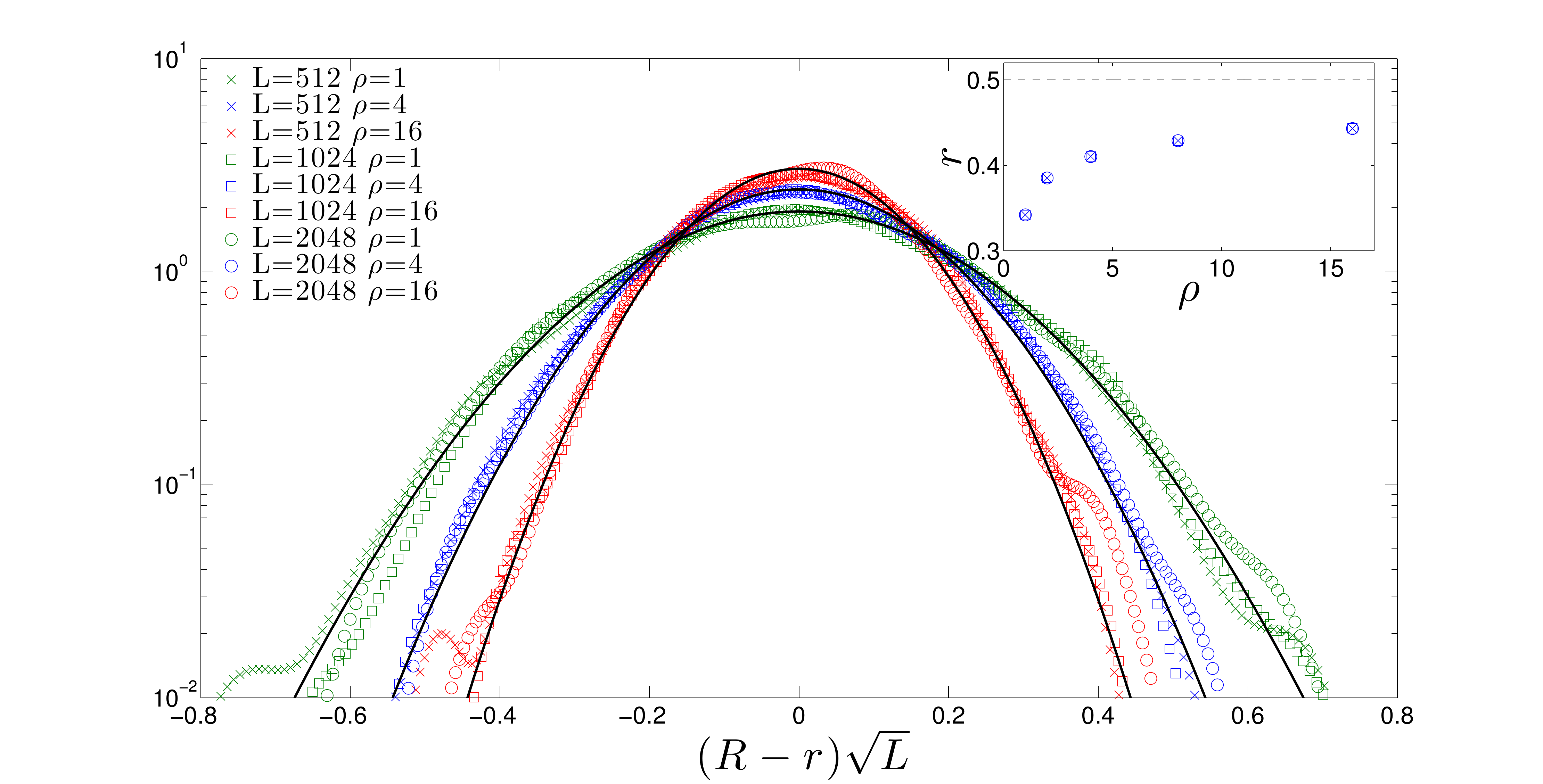}
\caption{Distribution of the ratio of occupied sites $R$ (\ref{Rdef}) (centred and scaled by $\sqrt{L}$) for TASIP. Black curves are probability density functions of Gaussians with mean $0$ and standard deviations from data sets $L=2048$ and corresponding $\rho$. The inset shows $r$ for systems with size $L=512,1024, 2048$ and density $\rho=1,2,4,8,16$. Data collapse confirms that $r$ depends only on $\rho$, and has an upper bound $0.5$ ($\rho\rightarrow\infty$). Distribution functions from data (2000 realisations) are a kernel density estimate computed by ksdensity with Matlab.}
\label{TASIPPdfR}
\end{center}
\end{figure}

As is shown in Figure~\ref{TASIPPdfR}, the ratio $R$ of occupied sites at the end of the nucleation regime follows a Gaussian distribution with density dependent fluctuations of order $1/\sqrt{L}$. The mean $r$ monotonically increases with $\rho$ from values around $0.35$ for small densities $\rho\approx 1$. This can be consistently explained using a toy model of coalescing particles, which is presented in appendix \ref{Appendix1}. For large densities alternating occupied/empty patterns are observed with long correlation lengths, and in the limit $\rho\to\infty$ we expect the system to approach the maximal theoretical value $r=0.5$. The inset in Figure~\ref{TASIPPdfR} shows that this convergence is slow and is an interesting question for further investigation. In this paper we focus on other aspects of the dynamics, and will use the actual value of $r$ as a fit parameter in the next sections.

\section{Condensate motion and interaction}\label{sec:cond}

In this section we analyse the motion of an isolated macroscopic cluster which dominates the stationary dynamics of the model. We further study the interaction between condensates, which is the foundation of understanding the coarsening and saturation dynamics as discussed in the next section. 

\subsection{Dynamics of isolated clusters}\label{sec:cond:isoclust}

\subsubsection*{Totally Asymmetric Dynamics}

Consider an isolated cluster of large size $m\gg1$ on site $x$, for simplicity on an otherwise empty lattice. The only possible transition is that a particle jumps to site $x+1$ with rate $d_{L} m$. Then the single particle could move to site $x+2$ at rate $d_{L}$, or the condensate could loose another particle with much higher rate $m-1$ due to the inclusion interaction. Thus, given that no particle exits to site $x+2$, the total time $T_{\textrm{step}}$ for all particles to move to site $x+1$ is a sum of independent exponential variables 
with mean
\begin{align}
\mathbb{E}[T_{\textrm{step}}]&=\frac{1}{d_L m} {+}\sum_{k=1}^{m-1}\frac{1}{(m-k)k} \cong\frac{1}{d_L m} {+}\frac{1}{m}\int_{1/m}^{(m-1)/m}\frac{1}{x(1 -x)}\, dx\\
&\cong \frac{1}{d_L m} {+}\frac{2}{m}\log(m) \, . \nonumber
\end{align}
Due to the quadratic scaling of the inclusion interaction the process speeds up significantly after the first particle and the remaining time vanishes with respect to the time of the initial move. In particular $T_{\textrm{step}}$ is dominated by the exponential time of the first particle, so to leading order $T_{\textrm{step}}\sim \mathrm{Exp}(d_L m)$. The rate at which any particle escapes from site $x+1$ is bounded above by $d_{L} m\rightarrow 0$ with $L\rightarrow \infty$. Thus, in the limit a macroscopic cluster is stable and jumps to the right with vanishing rate $d_L m\to 0$ which is proportional to its size. In general, the time scale for motion of macroscopic clusters of size $O(L)$ or the stationary single condensate is 
\be\label{tmovetasip}
\tau^{\textrm{move}}_{L}=\frac{1}{d_{L}L}\, .
\ee
This is consistent with results in \cite{Chleboun:2012vc} on the vanishing stationary current, which is dominated by the motion of a single condensate as
\be
j(\rho )\cong d_L \rho^2 L^2 /L= \rho^2 d_L L\ .
\ee

\subsubsection*{Symmetric Dynamics}
For symmetric dynamics a single cluster on an otherwise empty lattice is also stable, but performs a symmetric continuous-time random walk. Analogous to the above, the first particle from site $x$ moves with rate $d_{L}m$ to site $y=x-1$ or $x+1$. Then the inclusion interaction dominates the dynamics, and particles are exchanged symmetrically between sites $x$ and $y$ until one of them is again empty.  We find
\be
\mathbb{E}[T_{\textrm{step}}]\simeq \frac{1}{d_L m} +O(1) \, ,
\ee
since the expected number of steps is $m-1$ (\ref{excursion}) and the largest expected waiting time is $1/(m-1)$ (see Appendix \ref{Appendix2}). So the step is again dominated by the motion of the first particle. The jump attempt of the cluster is only successful if all particles end up on the new site $y$ rather than $x$, which happens with probability $1/m$ (\ref{p0}). So the cluster performs a symmetric random walk with effective rate $d_L$ and the time scale for cluster motion is 
\be\label{tmovesip}
\tau^{\textrm{move}}_{L}=\frac{1}{d_{L}}\,.
\ee

\subsection{Interaction of Two Clusters}\label{sec:cond:twoclust}
\subsubsection*{Totally Asymmetric Dynamics}
In TASIP, we have seen above that isolated clusters jump to the right with rate proportional to their speeds. Therefore, they move freely until a faster cluster catches up with a smaller one. As soon as they are only one intermediate lattice site apart they start interchanging particles via a mechanism first observed in \cite{Waclaw:2012ww}. To describe this situation let at time $0$, $\eta_1 >\eta_3$, both of order $m\gg1$ and $\eta_2 =0$ on the intermediate site. Then it is more likely that site $1$ looses a particle to $2$ rather than site $3$ to $4$ and the clusters start interacting. The dynamics is then dominated by the inclusion rate and we can ignore the diffusion part. Ignoring jumps away from site $3$, the only two events are jumps from site $1$ to $2$ or from site $2$ to $3$ with rates of order $\eta_1 (t)\eta_2 (t)$ and $\eta_2 (t)\eta_3 (t)$, respectively. Therefore, the probabilities for the next event to be of type one or two are
\be\label{jprobs}
\frac{\eta_1 (t)}{\eta_1 (t)+\eta_3 (t)}  \quad\mbox{and}\quad \frac{\eta_3 (t)}{\eta_1 (t)+\eta_3 (t)}\ .
\ee
The interaction process continues on the simplex $\eta_1 (t),\eta_3 (t)\leq \eta_1 +\eta_3$ following left-up paths due to total asymmetry, until the cluster on site $3$ moves to site $4$ which becomes more likely once $\eta_3 (t)>\eta_1 (t)$ and $\eta_{2}(t)=0$. Note that the result of the mass redistribution depends only on the discrete embedded chain with probabilities (\ref{jprobs}), which exhibit a symmetry under exchanging sites $1$ and $3$ with invariant diagonal $\eta_1 =\eta_3$. Since the whole process is invariant under time and space inversion, the statistics of all paths leading towards the diagonal for $\eta_1 >\eta_3$ is the same as that of all paths leading away. The cluster interaction is therefore symmetric, such that in distribution $\eta_1 (T)\stackrel{dist}{=}\eta_3(0)$ and $\eta_3 (T)\stackrel{dist}{=}\eta_1(0)$ where $T$ is the time when the first particle moves from site $3$ to $4$ and the interaction terminates. So to leading order the clusters penetrate each other and just exchange places, and along the way exchange an unbiased amount of $O(\sqrt{m})$ particles due to fluctuations.

Note that the above description is qualitative but exact, and can also be corroborated by the solution to a scaling limit of the standard evolution equations for the rescaled masses $\rho_x =\eta_x /N$. We consider the situation in which all $N$ particles in the system reside on 3 sites $x=1,2,3$, i.e. $\rho_1 +\rho_2 +\rho_3 =1$ and $\rho_{x}=0$ otherwise. Now consider the rescaled process $\left(\bm{\rho}(t) \,: \,t\geq0\right)$ defined by
\bee
\bm{\rho}(t):=\left(\eta_{x}(t)/N \,:\,  x\in \{1,2,3 \} \right) \, .
\eee
This is a Markov process on the simplex $E=\left\{[0,1]^{3},\sum_{x=1,2,3}\rho_{x}=1\right\}$ with generator

\be\label{TEGe}
\mathcal{L}_{N}f(\bm{\rho})=\sum_{x=1,2}N\rho_{x}(d_{N}+N\rho_{x+1})\left(f(\bm{\rho}-\frac{1}{N}\bm{e}_{x}+\frac{1}{N}\bm{e}_{x+1})-f(\bm{\rho})\right)\, ,
\ee
where we can again ignore any particle leaving to site $4$. Initially, a small initial mass is on site 2: $\rho_{2}=\epsilon=1-\rho_{1}-\rho_{3}= O(1/N)\ll 1$. Then assuming $f$ is smooth, Taylor expansion of right hand side  gives 
\bee
\mathcal{L}_{N}f(\bm{\rho}) {=}\!\!\! \sum_{x=1,2}\!\!\!N\rho_{i}(d_{N}{+}N\rho_{i+1})\!\!\left[\!\left(\!\frac{1}{
N}\!\!\left(\partial_{\rho_{i+1}}{-}\partial_{\rho_{i}}\right)\!{+}\frac{1}{2N^{2}}\!\left(\partial_{\rho_{i+1}}{-}\partial_{\rho_{i}}\right)^{2}\right)\!\!f(\bm{\rho}){+}O(\frac{1}{N^{3}})\right] \, ,
\eee
where we use abbreviation 
\bee
\left(\partial_{\rho_{i}}-\partial_{\rho_{j}}\right)^{2}=\frac{\partial^{2}}{\partial{\rho_{i}}^{2}}-2\frac{\partial^{2}}{\partial\rho_{i}\partial\rho_{j}}+\frac{\partial^{2}}{\partial\rho_{j}^{2}} \, .
\eee
For large systems $d_{N}$ terms are negligible and for the test function$f(\bm{\rho})=(\rho_{1},\rho_{3})$ we get,
\be\label{ScalTFGe}
\frac{1}{N}\mathcal{L}_{N}\begin{pmatrix}\rho_{1}\\\rho_{3}\end{pmatrix}=-\rho_{1}\rho_{2}\begin{pmatrix}1\\0 \end{pmatrix}+\rho_{2}\rho_{3}\begin{pmatrix}0\\1\end{pmatrix}+O(\frac{1}{N}) \, .
\ee
Note that to leading order the second order derivative terms cancel, so $\rho_{i}(t)$ is  deterministic, and the order of the fluctuation terms are consistent with the unbiased exchange of order $\sqrt{N}$ particles as discussed above. Ignoring the correction term and slowing down time by taking $t\mapsto t/N$, with (\ref{ScalTFGe}) the evolution equation gives

\bee
\frac{d}{dt}\begin{pmatrix}\rho_{1}(t)\\ \rho_{2}(t)\end{pmatrix}=\mathcal{L}\begin{pmatrix}\rho_{1}(t)\\ \rho_{2}(t)
\end{pmatrix}=\begin{pmatrix}-\rho_{1}(t)\rho_{2}(t) \\  \rho_{2}(t)\rho_{3}(t)
\end{pmatrix} \, ,
\eee
where we used $\mathbb{E}[\rho_{i}]=\rho_{i}$ and $ \rho_{2}=1-\rho_{1}-\rho_{3}$. For initial conditions $\rho_{1}(0)$ and $\rho_{3}(0)$ such that $(\rho_{1}(0)+\rho_{3}(0))<1$, $2\rho_{1}(0)>1$, we have the solution: 
\begin{align*}
\rho_{1}(t)&=\frac{1}{2}\left(1-D\tanh\left(\frac{Dt}{2}-A\right)\right)\rightarrow \frac{1-D}{2} \,\,\, \textrm{as} \,\,\, t\rightarrow \infty  \, ,\\
\rho_{2}(t) &=\frac{2\rho_{1}(0)\rho_{3}(0)}{1-D\tanh\left(\frac{Dt}{2}-A\right)} \rightarrow \frac{2\rho_{1}(0)\rho_{3}(0)}{1-D}  \,\,\, \textrm{as} \,\,\, t\rightarrow \infty \, ,\\
\textrm{where} \,\,\, D&=\sqrt{1-4\rho_{1}(0)\rho_{3}(0)} \,\,\,\textrm{and}\,\,\, A=\tanh^{-1}\left(\frac{2\rho_{1}(0)-1}{D}\right) \nonumber \,\,\,.
\end{align*}
We have $D=\sqrt{1-4\rho_{1}(0)(1-\epsilon-\rho_{1}(0))}\rightarrow (2\rho_{1}(0)-1)>0$ as $\epsilon\to 0$, which implies $\rho_{1}(t)\to\rho_{3}(0)$ and $\rho_{3}(t)\to\rho_{1}(0)$ as $t\to\infty$, and the clusters exchange places.

\subsubsection*{Symmetric Dynamics}
For symmetric dynamics, the mechanism of cluster interaction is different and has been established in \cite{Grosskinsky:2013ji}. Two clusters on next-nearest neighbour sites, say $1$ and $3$, of rescaled sizes $\rho_1 ,\rho_3 \in[0,1]$ with initially $\rho_1 =\rho_3 =1$ can continuously exchange mass on the slow time scale $d_L$ via the intermediate site according to the Wright-Fisher-type generator $\rho_{1}\rho_{3}(\partial_{\rho_{3}}-\partial_{\rho_{1}})^{2}$, which conserves the total mass. In addition, both clusters can merge on site $2$ in a jump event with rate $\rho_1 +\rho_3$. Since both clusters can separate only with site $1$ moving to the left with rate $\rho_1$ and site $3$ to the right with rate $\rho_{3}$, the coalescence event actually happens with probability $1/2$. But even without merging, the continuous exchange will lead to a finite fraction of particles being redistributed in an unbiased fashion, so that in a typical interaction event of order $N$ particles are exchanged, in contrast to $\sqrt{N}$ for totally asymmetric dynamics.

\subsection{Derivation of time scale}\label{sec:cond:timescale}

The mechanism of cluster interaction together with the time scale for motion $\tau_{L}^\textrm{{move}}$ (\ref{tmovetasip}) and (\ref{tmovesip}) determines the the time scale $\tau_{L}$ of coarsening and relaxation of the system, which we used in Figure \ref{TASIP2ndMoment}. For the TASIP, condensates containing of order $\rho L$ particles have speed of order $d_{L}\rho L$. Then the relative speed between any two condensates is also of this order, which leads to the average time between two encounters to be of order $L\cdot \tau_{L}^{\textrm{move}} \sim 1/(\rho d_{L})$. Since every interaction leads to an unbiased exchange of order $\sqrt{\rho L}$ particles, order $\rho L$ exchanges are necessary to achieve a macroscopic change, which leads to the time scale
\be
\tau_{L}^{a}=L/d_{L} \ ,
\label{TimeScaleTASIP}
\ee
which is independent of the particle density $\rho$.

Following the similar argument for the SIP, the average time between successive encounters becomes $L^{2}\cdot\tau_{L}^{\textrm{move}} \sim L^{2}/\rho d_{L}$, since the condensates need to perform order $L^{2}$ jumps to meet as they perform symmetric random walks with rate $1/(\rho d_{L})$. But as opposed to the TASIP, condensates can exchange a macroscopic amount of particles so that we only need $O(\rho )$ such encounters, leading to
\be
\tau_{L}^{s}=L^2/d_{L} \ ,
\label{TimeScaleSIP}
\ee
which is again independent of $\rho$.

\section{Coarsening and saturation}\label{sec:cors}

\subsection{Dynamics in the coarsening regime}\label{sec:cors:dym}

We use heuristic arguments to derive the coarsening dynamics, based on the dynamics of a single `typical' cluster and its interaction with others in a mean-field approximation.

\subsubsection*{Totally Asymmetric Dynamics}

Let $m(t)$ denote the typical size of a cluster in the coarsening regime, and $n(t)$ the typical number of clusters per volume, so that we have $n(t)\, m(t)=\rho$. 
We denote the speed of a typical cluster by $v(t)=d_{L}m(t)$ and the typical distance of two clusters is given by $s(t)=m(t)/\rho$. Then the rate at which two clusters meet is $v(t)/s(t) = \rho d_{L}$. As discussed in Section \ref{sec:cond:twoclust}, when two clusters meet, they make an unbiased exchange of order $\sqrt{m}$ particles. So for one cluster to lose all its particles, it typically takes of order $m(t)$ exchanges. Therefore, each cluster independently disappears with rate $C_{a}\rho d_{L}/m(t)$, where $C_{a}$ is a proportionality constant which is hard to predict and we will just fit it from simulation data. These death events, which happen typically after time $\Delta t=m(t)/(C_{a}\rho d_{L} n(t))$ per unit volume, drive the coarsening process. Each event effectively increases $m(t)$ by $\Delta m(t)=m(t)/n(t)$ per unit volume, which leads to
\be
\frac{d}{dt}m(t) = \frac{\Delta m(t)}{\Delta t}=
C_a \rho d_{L}\, .
\label{CorsODE}
\ee
The initial condition is 

\begin{figure} 
  \subfigure[]{ 
    \label{TASIPCors} 
    \begin{minipage}[b]{0.5\textwidth} 
      \centering 
      \includegraphics[height=5cm]{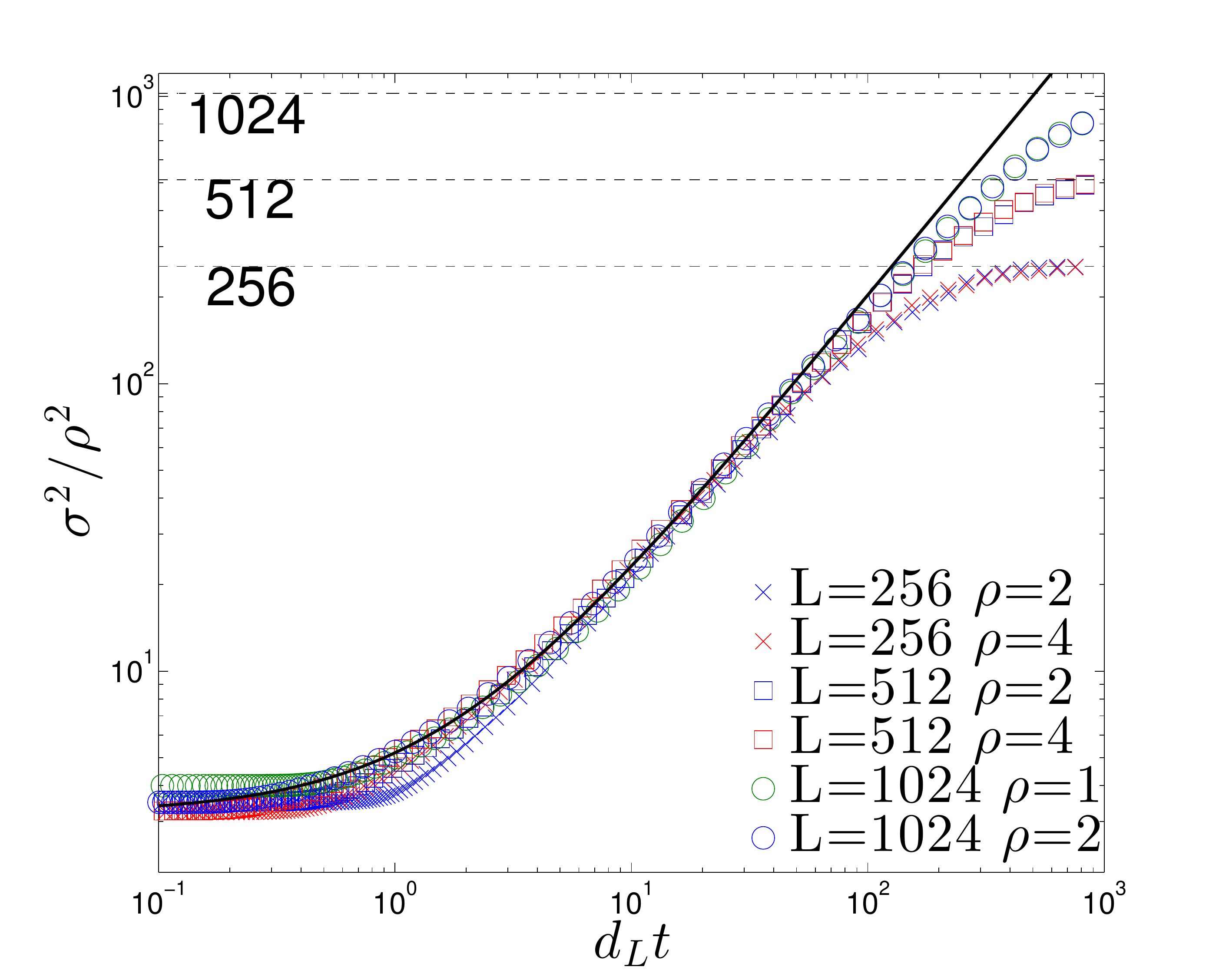} 
    \end{minipage}}%
  \subfigure[]{ 
    \label{SIPCors} 
    \begin{minipage}[b]{0.5\textwidth} 
      \centering 
      \includegraphics[height=5cm]{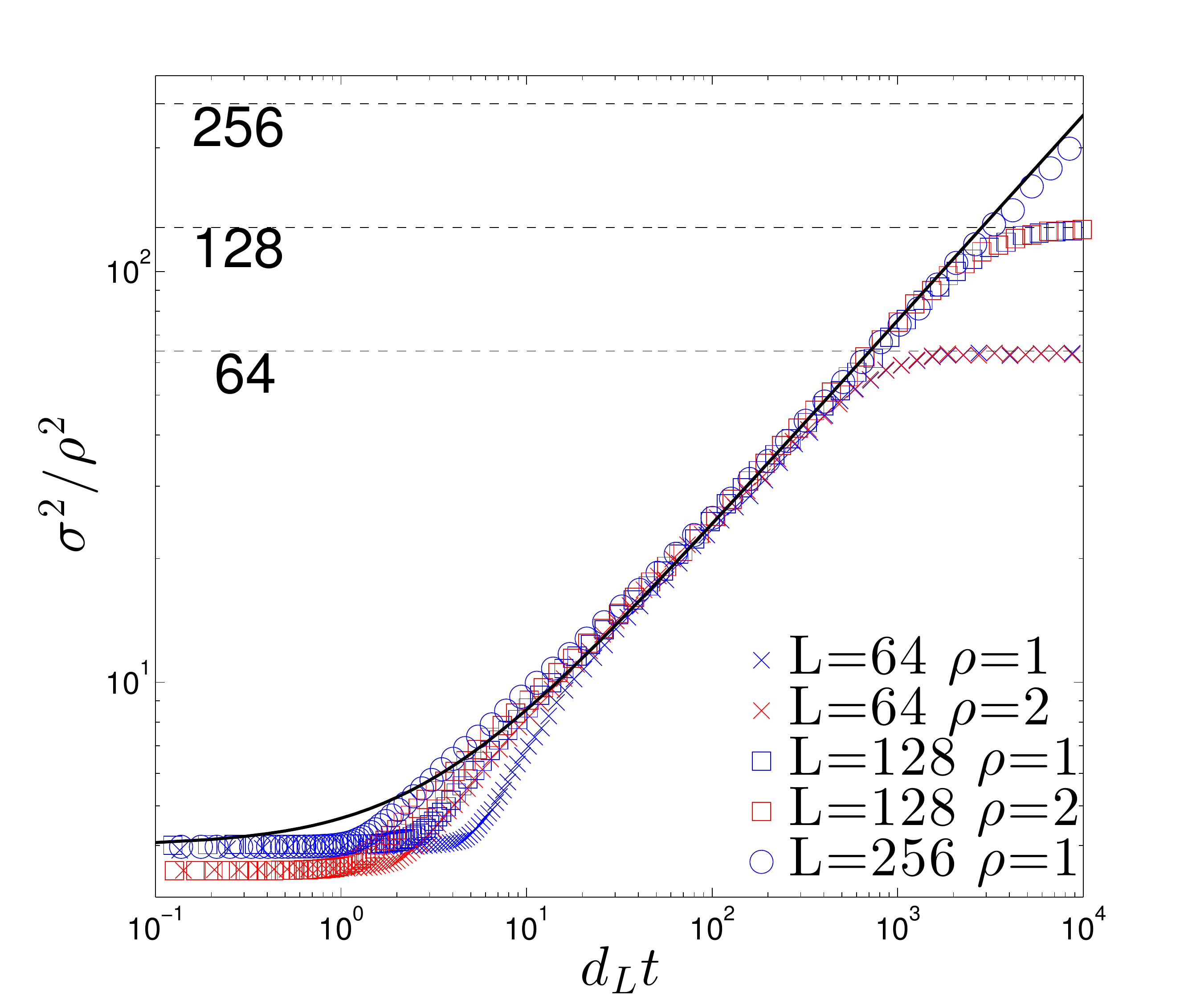} 
    \end{minipage}} 
  \caption{Power-law scaling of $\sigma^{2}(t)/\rho^{2}$ (\ref{smom}) in the coarsening regime. (a) Data for TASIP compared to the prediction (\ref{PwTASIP}) shown as a full line with fitted constant $C_{a}=1.8961$ and initial condition $r=0.3851$. (b) Data for SIP compared to the prediction (\ref{PwSIP}) shown as a full line with fitted constant $C_{s}=5.7614$. Data points are averaged over 200 realisations. Errors are bounded by the size of the symbols.} 
  \label{Cors} 
\end{figure}

\bee
m(0)=\frac{\rho }{n(0)}= \frac{\rho}{r} \ ,
\eee 
where $n(0)=r$ is the expected ratio of occupied sites after nucleation $r$ (\ref{Rdef}) which we also fit from the data. The solution to (\ref{CorsODE}) is then simply
\be
m(t)=C_a \rho d_{L}t+\frac{\rho}{r}.
\ee
Due to the clustered nature of configurations during the coarsening regime we have
\bee
\sigma^{2}(t)=\frac{m^{2}(t)}{s(t)}=\rho\, m(t)=C_a \rho^{2}d_{L}t+\frac{\rho^{2}}{r}\ ,
\eee 
which implies
\be\label{PwTASIP}
\frac{\sigma^{2}(t)}{\rho^{2}}=C_a d_{L}t+\frac{1}{r}\ .
\ee
Note that there is no explicit system size dependence in the above analysis and this scaling law also holds on infinite lattices (given a fixed small parameter $d_{L}$). On a finite lattice it only applies in a certain scaling window, after which the system saturates due to finite size effects (see Figure \ref{Cors}(a)), reminiscent of the classical Family-Viscek scaling for coarsening dynamics in surface growth (see e.g. Chapter 3.3 in\cite{Family:1991uz}).
The time scale $\tau_{L}$ characterises this scaling window and the relaxation of the system, and is determined by the scaling solution reaching its maximal stationary value, i.e. 
\bee
m(\tau^a_{L})=C_a \rho d_{L}\tau^a_{L}+\frac{\rho}{r}=O(N)\ .
\eee
This implies $\tau^a_{L}=O (L/d_{L})$ and corresponds to the time when all clusters have merged to a single condensate. This agrees with our previous prediction for the asymmetric time scale in (\ref{TimeScaleTASIP}).

\FloatBarrier

\subsubsection*{Symmetric dynamics}
We can apply the same argument to the SIP and get similar results. Since particles jump symmetrically, the velocity of clusters is replaced by the diffusivity $D=d_{L}$ (\ref{tmovesip}), and the interaction rate is then $D/s^{2}$. Unlike the TASIP, a single interaction of two clusters in the SIP leads to macroscopic exchange of order $m(t)$ particles as was derived in Section \ref{sec:cond:twoclust}. Then we have

\be\label{syeq}
\frac{d}{dt}m(t)=C_s \frac{Dm(t)}{s^{2}}=C_s \frac{d_{L}\rho^{2}}{m(t)} \,\,\, ,
\ee
where $C_s $ is again a proportionality constant for cluster interaction. With initial condition $m(0)$, we have the solution
\bee
m(t)=\sqrt{2C_s \rho^{2}d_{L}t+m(0)^2}\, .
\eee
As before, the second moment can be written as 
\bee
\sigma^{2}(t)=\rho m(t)=\rho^{2}\sqrt{2C_s d_{L}t+(\sigma^2 (0)/\rho^2 )^2}, 
\eee
and for the initial condition we now have the exact result of the nucleation regime (\ref{NRSIPSecMotLmt}) where $\sigma^{2}(0)/\rho^{2}=3+1/\rho$. This leads to
\be\label{PwSIP}
\frac{\sigma^{2}(t)}{\rho^{2}}=\sqrt{2C_s d_{L}t+\left(3+\frac{1}{\rho}\right)^{2}} \, ,
\ee
where we only have to fit the parameter $C_{s}$. 
This scaling law is confirmed in Figure\ref{Cors}(b), and the scaling window and time scale can again be determined from
\bee
m(\tau^s_{L})=\sqrt{2C_s \rho^{2}d_{L}\tau^s_{L}+\frac{\rho^{2}}{r^{2}}}=O(N)\ .
\eee
This implies $\tau^s_{L}=O(L^{2}/d_{L})$ which agrees with our previous prediction in (\ref{TimeScaleSIP}).

\subsection{Exponential saturation and stationarity}\label{sec:cors:ex}

Having identified the time scales $\tau_L$ of the coarsening window for symmetric and asymmetric dynamics, we expect that the power-law behaviour turns into an exponential saturation of the system to the stationary value $1$ of our observable $\sigma^2 (t)/(\rho^2 L)$, i.e.
\be\label{exlaw}
\frac{\sigma^{2} (t)}{\rho^{2}L} \simeq 1-e^{-C' t/\tau_L} \quad\mbox{as }t\to\infty\ .
\ee
This is essentially equivalent to the assertion that $C'/\tau_L$ is indeed the spectral gap of the generator of the system, which usually describes the exponential approach to equilibrium in finite systems as described above.

\begin{figure} 
\centering
  \subfigure[TASIP]{
    \begin{minipage}[b]{0.5\textwidth} 
      \centering 
      \includegraphics[height=5cm]{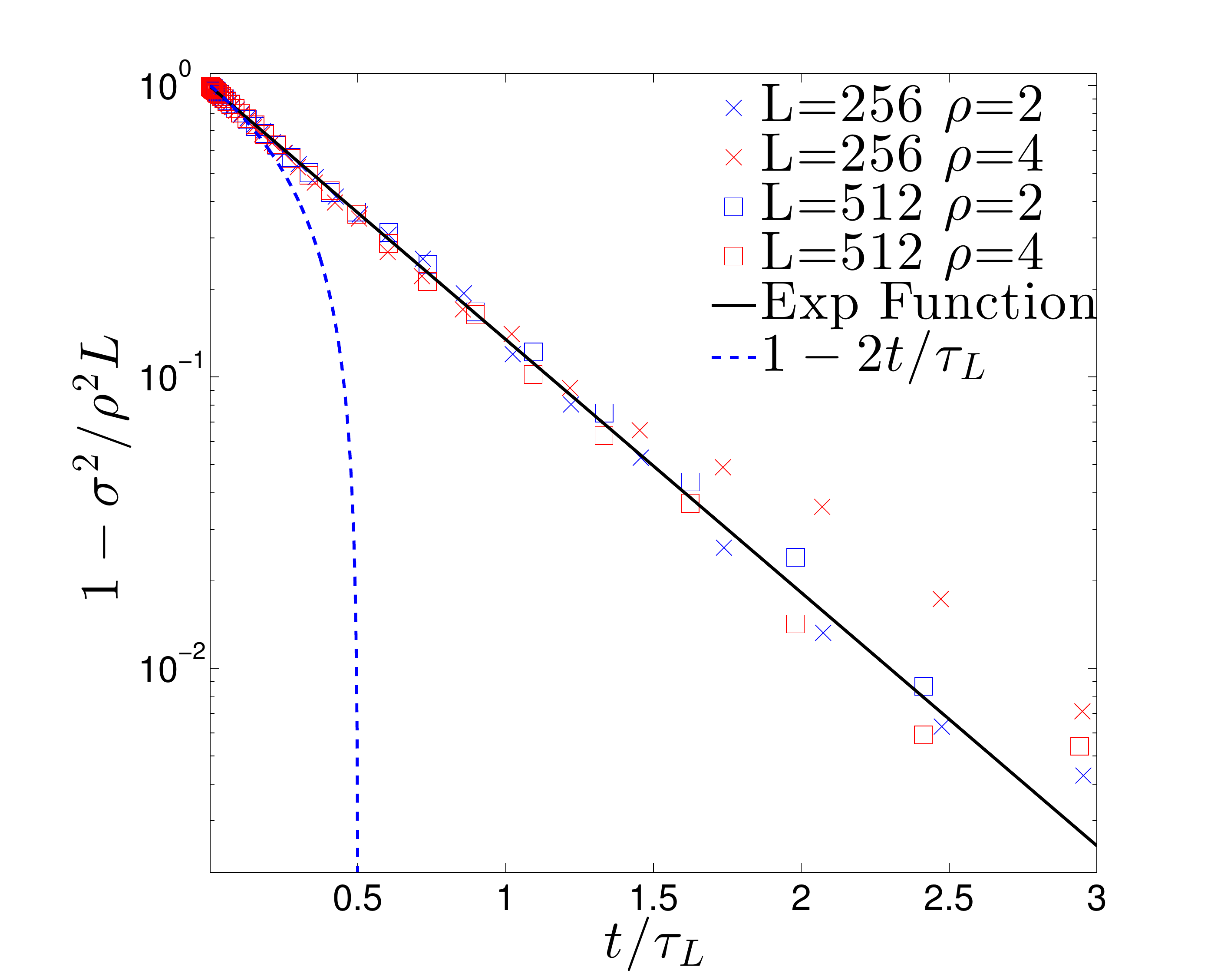} 
    \end{minipage}
    }%
  \subfigure[SIP]{ 
    \begin{minipage}[b]{0.5\textwidth} 
      \centering 
      \includegraphics[height=5cm]{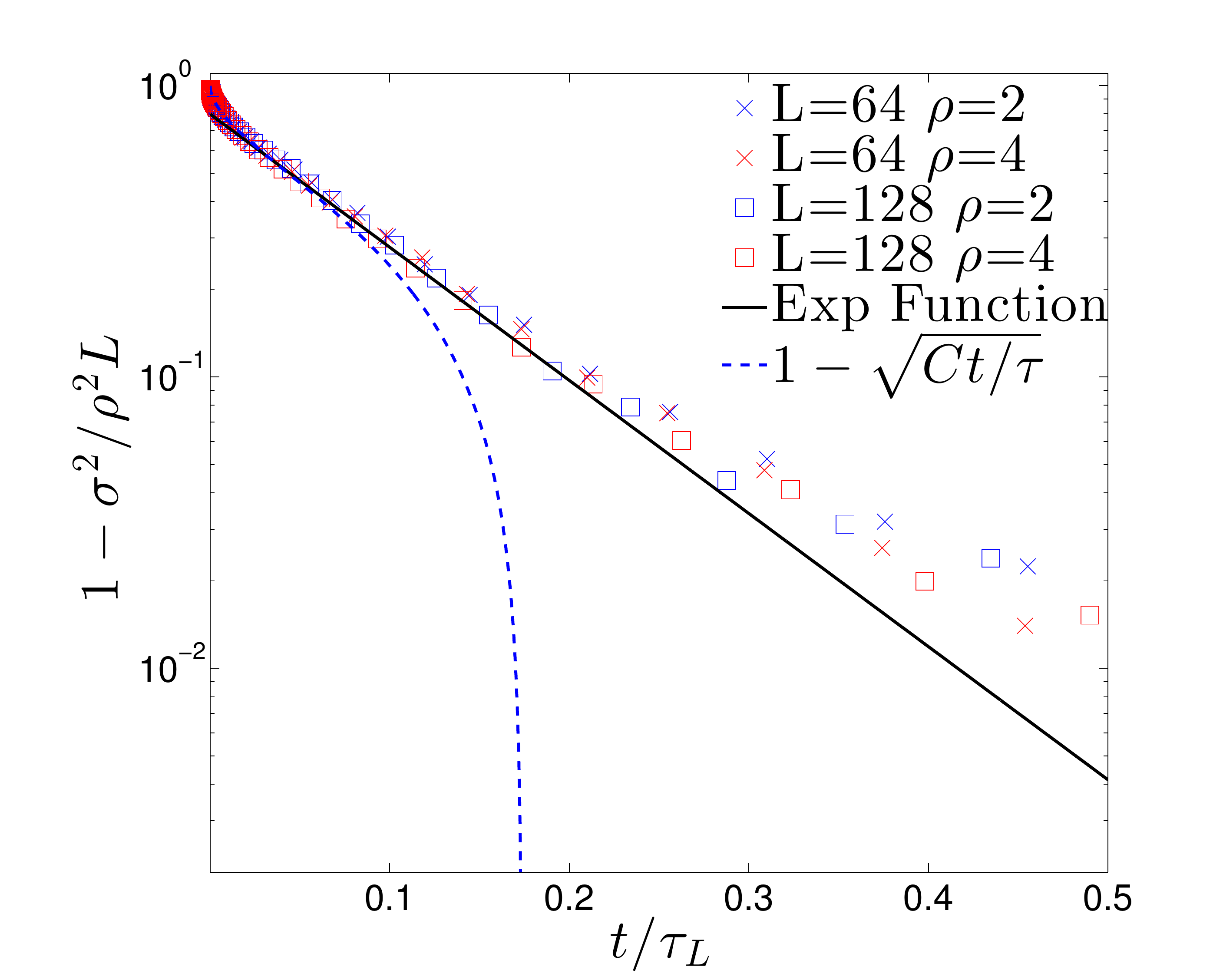} 
    \end{minipage}
    } 
  \caption{
	Exponential relaxation in the saturation regime for TASIP (a) and SIP (b). The predictions (\ref{exlaw}) are shown as a full lines with best fit constants $C'_a =2.00$ and $C'_s =10.51$. 
	In both cases we plot the coarsening scaling law (dashed line) for comparison, which is only valid for short times on the scale $\tau_L$. Data points are averaged over 200 realisations. Errors are of the order $10^{-4}$.}
  \label{ExpDecay} 
\end{figure}

For symmetric dynamics, we can provide a simple derivation which includes a rough estimate of the constant $C'_s$. The late stage of the dynamics is dominated by $2$ remaining clusters competing for particles. On average, both of them have roughly size $m\approx N/2$, and from the arguments in the previous subsection (\ref{syeq}) we see that under that assumption they meet at rate
\bee
C_s \frac{D}{s^2} =C_s \frac{4 d_L}{L^2} =4C_s /\tau_L^s \quad\mbox{since}\quad s=m/\rho =L/2\ .
\eee
As mentioned in Section \ref{sec:cond:timescale}, at each encounter the clusters can merge with probability $1/2$, which would lead to a single condensate and remaining in a typical stationary configuration. Since merge attempts are independent, this leads to an effective rate to reach stationarity roughly given by $2C_s /\tau_L^s$, and we expect
\be
1-\frac{\sigma^{2} (t)}{\rho^{2}L} \simeq e^{-C'_s t/\tau_L^s } \quad\mbox{as }t\to\infty\ ,
\ee
with $C'_s \approx 2C_s$. This is confirmed in Figure \ref{ExpDecay}(b), where we see a good data collapse with exponential decay with a best fit parameter $C'_s =10.51$, which is similar to $2C_s$ as fitted in Figure \ref{Cors}. Given the crude approximation of two equal sized clusters in our derivation we cannot expect a perfect match of those constants.

For totally asymmetric dynamics two macroscopic clusters cannot merge in a single interaction event, but exchange only of order $\sqrt{L}$ particles in an unbiased fashion. Still, we expect the approach to stationarity to be governed by an exponential law of the form (\ref{exlaw}), and we can derive the constant by direct comparison with the coarsening dynamics. Expanding (\ref{exlaw}) for times $t\ll \tau_L$ we get
\bee
\frac{\sigma^{2} (t)}{\rho^{2}L} \simeq 1-C'_a t/\tau_L^a =C'_a \frac{t\, d_L}{L} \, ,
\eee
where we used $\tau_L^a =L/d_L$. This matches the scaling law solution (\ref{PwTASIP}) and we see that in fact $C'_a =C_a$. Again this is confirmed in Figure \ref{ExpDecay}(a), where the best fit parameter for $C'_a$ is very close to $2$ as is $C_a$. We currently do not have a good theoretical explanation to predict this value, but our numerics strongly suggest that the constant in the asymmetric case seems to be simply $2$.

Note that the expansion of the exponential law matches with the coarsening law only for the totally asymmetric case, since the coarsening law (\ref{PwTASIP}) is in fact linear. This leads to the fact that the whole coarsening and saturation dynamics are well described by the exponential law, as can be seen in Figure \ref{TASIP2ndMoment}. For symmetric dynamics this matching argument would not work, since the coarsening law has exponent $1/2$ and the exponential approximation is simply not valid in the coarsening window.

For large values of $t$ the deviation from the exponential decay in Figure \ref{ExpDecay} is again a finite size effect. The stationary value of $\sigma^{2} /(\rho^{2}L)$ is slightly smaller than $1$, due to the fact that the single condensate continues to move on the time scale $\tau_L$. During a step the mass is temporarily distributed on two lattice sites, which decreases the stationary average of $\sigma^{2}$. We have estimated a similar contribution to nearest-neighbor correlation functions in Section \ref{sec:nucl} using an ergodic average, and an analogous computation leads to stationary corrections of the order $1-\sigma^{2} /(\rho^{2}L) \sim d_L /\rho$ for symmetric and totally asymmetric dynamics.

\section{Conclusion}\label{sec:cons}

We have derived a heuristic description of the dynamics of condensation of the asymmetric inclusion process in the thermodynamic limit. We identified three dynamical regimes, and the main focus was the derivation of a coarsening scaling law. Our predictions have been confirmed by extensive simulations and describe the actual dynamics very well, in particular in the totally asymmetric case. Our arguments are based on the analysis of the dynamics of a typical cluster and interaction with others in a mean-field approximation, which is justified by observation of typical time evolutions of the system. This approach does not work for the explosive condensation model studied in \cite{Waclaw:2012ww}, where the full dynamics is dominated by a single large cluster and leads to a relaxation time scale that is decreasing with the system size.

The symmetric dynamics have been included mostly for comparison and to better understand the complicated dynamics for the totally asymmetric case in the nucleation regime. Since the symmetric inclusion process is self-dual, time dependent correlation functions can be computed exactly, which we have used indirectly for the nucleation regime. This holds, however, for the whole dynamics of the process, and a more detailed analysis of the duality structure of the process is expected to lead to a rigorous description of the full time evolution in the thermodynamic limit. This is current work in progress. A further interesting question arising for future work is a better understanding of the dynamics of the nucleation regime in the asymmetric case.

\section*{Acknowledgements}
J.C. acknowledges funding from the University of Warwick through a Chancellor's International Scholarship. S.G. acknowledges support by the Engineering and Physical Sciences Research Council (EPSRC), Grant No. EP/I014799/1. P.C. acknowledges support and funding from the University of Warwick, Institute of Advanced Study through a Global Research Fellowship. We are grateful for inspiring discussions with F.~Redig and M.R.~Evans.

\bibliographystyle{unsrt}
\bibliography{inclu}
\begin{appendix}

\section*{Appendix}
\setcounter{section}{1}
\setcounter{subsection}{0}

\subsection{Toy model for the nucleation regime}\label{Appendix1}
We define a toy model for the number of occupied sites after the nucleation regime of the TASIP on the lattice $\Lambda_L=\{1,2,3,...,L\}$ with periodic boundary conditions, where the modified state variable $\eta_x \in\{ 0,1\}$ simply indicates whether site $x$ is occupied. We consider the simplest uniform initial distribution $\eta_{x}(0)=1$ for all $x\in\Lambda_{L}$. After waiting time $T_{x}$, the mass on site $x$ tries to jump to site $x+1$, where $T_{x}$ are i.i.d. random variables. This jump is successful only if $\eta_{x+1} (T_x )=1$, i.e. the mass on site $x+1$ has not moved before, and after the jump we have $\eta_x =0$ and $\eta_{x+1} =1$. This is a simplified model of the inclusion interaction of the process in the nucleation regime, and keeps track only of occupied sites. The $T_x$ can be interpreted as the random times when the full mass in the true TASIP has moved from site $x$ to $x+1$. The distribution of those times is not important for our argument, we only assume that they are independent, and their order therefore corresponds to a uniform permutation.

After some time all particles reside on non-successive sites and the toy model reaches an absorbing state. Such absorbing configurations are constructed by blocks in different lengths, where one block has several empty sites and only one occupied site on the rightmost site of this block. In other words, the blocks are of the form $000\ldots 001$. We denote the length of such a block (indexed by $n$) by $X_{n}\in\N$, where $2\leq X_{n}\leq L$ and $\sum_{n}X_{n}=L$. 
Assume that the occupied site of one such block is $z$, then $\eta_{z-1}=\eta_{z-2}=...=\eta_{z-k}=0$ when $k+1$ is the size of the block. The event $X_{n}\geq k+1$, i.e. a block size of at least $k+1$, is equivalent to the event 
\be
T_{z-k}<T_{z-k+1}<\ldots <T_{z-1} \,\,\, ,
\label{ts}
\ee
since each initial particle has to jump earlier than its right neighbour, so that all the mass on these sites could move up to site $z$. 
Since the times are ordered in a uniform permutation, the probability for (\ref{ts}) determines
\bee
\mathbb{P}(X_{n}-1\geq k)=\frac{1}{k!} \, .
\eee 
So we get the following limiting behaviour for the expectation,
\bee
\mathbb{E}[X_{n}]=\sum_{k=1}^{L-1}\frac{1}{k!}+1\rightarrow e \,\,\, , \,\,\, \textrm{as} \,\,\, L\rightarrow \infty \,\,\, .
\eee
Note that the lengths of successive clusters are independent, so that the $(X_n :n\geq 0)$ constitute a renewal process on $\Lambda_L$, and
\bee
n_L :=\max\left\{n\,\,: \,\, \sum_{i=1}^{n}X_{i} \leq L\right\}
\eee
is the number of blocks at the absorbing state, which is equal to the number of remaining particles. From the standard renewal theorem (see e.g. Chapter 10.2 in \cite{Grimmett:2001wi}) we get
\bee
\frac{n_L}{L} \rightarrow \frac{1}{\mu} \quad\textrm{as}\,\,\, L\rightarrow\infty \,\,\,\textrm{almost surely}\,\,\, ,
\eee
where $\mu=\mathbb{E}[X_{1}]=e$ is the expected block length. Therefore, we have an approximation of the ratio of occupied sites (see (\ref{Rdef}))
\bee
r\approx 1/e =0.368\ .
\eee
This is very close to the observed value in Section \ref{sec:nucl} for small densities $\rho\approx 1$, where we expect the toy model to give the best approximation. For very low densities $r$ is dominated by the initial number of empty sites, whereas for high densities correlations built up over long distances leading to striped patterns, and $r$ seems to grow slowly towards its maximal value $1/2$ as $\rho\to\infty$.\\

\subsection{Symmetric random walk with absorbing boundary}\label{Appendix2}

Consider a simple symmetric discrete-time random walk $\{S_{n}, n\in\N\}$ with state space $X_{m}=\{0,1,2,...,m\}$ such that $S_{n}\in\{0,m\}$ are two absorbing states, and we define the excursion length $T:=\min\{n\in\N: S_{n}\in\{0,m\}\}$.
One can easily check that $S_{n}$ is a martingale, i.e. $\mathbb{E}[S_{n}|S_{0}]=S_{0}$, and then we have by the optional stopping theorem (see e.g. Chapter 12.5 in \cite{Grimmett:2001wi})
\bee
\mathbb{E}[S_{n}|S_{0}]=S_{0}=p_{0}\cdot0+(1-p_{0})\cdot m \, ,
\eee
where $p_{0}$ is the probability of the walker being absorbed in site $0$. Assume $S_{0}=k$, $k\in X_{m}$, then $\mathbb{E}[S_{0}]=k$ gives 
\be\label{p0}
p_{0}=1-k/m \, .
\ee
Define a new process $\{Y_{n}:=S_{n}^{2}-n, n\in\N\}$ which is also a martingale, since
\bee
\mathbb{E}[Y_{n+1}|Y_{n}]=\frac{1}{2}\left((S_{n}+1)^{2}-(n+1)+(S_{n}-1)-(n+1)\right)=S_{n}^{2}-n=Y_{n} \,.
\eee
Again, the martingale property and optional stopping theorem imply
\bee
\mathbb{E}[Y_{n}|Y_{0} =k]=k^{2} =-p_0 \mathbb{E}[T|Y_{0} =k] +(1-p_0 )\big( m^2 -\mathbb{E}[T|Y_{0} =k]\big)\ ,
\eee
and therefore 
\be\label{excursion}
\mathbb{E}[T|Y_0 =k]=(1-p_{0})m^{2}-k^{2}=k(m-k)\ .
\ee

\end{appendix}

\end{document}